\documentclass[aps,pra,twocolumn,showpacs]{revtex4}
\usepackage{amsmath,amssymb,mathptmx}
\DeclareSymbolFont{symbols}{OMS}{cmsy}{m}{n}
\renewcommand{\leq}{\leqslant}

\begin{document}

\title{Kraus representation of destruction of states for one qudit}

\author{Pawe{\l} Caban}
\email{P.Caban@merlin.fic.uni.lodz.pl}
\author{Kordian A. Smoli{\'n}ski}
\email{K.A.Smolinski@merlin.fic.uni.lodz.pl}
\author{Zbigniew Walczak}
\email{Z.Walczak@merlin.fic.uni.lodz.pl}

\affiliation{Department of Theoretical Physics, University of
  {\L}{\'o}d{\'z}, Pomorska 149/153, 90-236 {\L}{\'o}d{\'z}, Poland} 

\date{29 September 2003}

\begin{abstract}
  Quantum operations arise naturally in many fields of quantum
  information theory and quantum computing.  One of the simplest
  example of quantum operation is the von Neumann{\textendash}L{\"u}ders measurement.
  Destruction of states in quantum mechanics can be treated as a
  supplement to the von Neumann{\textendash}L{\"u}ders measurement [P.~Caban,
  J.~Rembieli{\'n}ski, K.~A. Smoli{\'n}ski and Z.~Walczak, J.\ Phys.\ A
  \textbf{35}, 3265 (2002)].  We show that destruction of states in
  one-qudit system is a quantum operation by finding its Kraus
  representation.
\end{abstract}


\maketitle

A general state change in quantum mechanics is described by the
quantum operations formalism \cite{Kra83} (see also \cite{Nie00} and
references therein) arising naturally in many fields of quantum
information theory and quantum computing.  In this formalism an input
state $\rho$ of a system is connected with an output state
$\mathcal{E}(\rho)$ by a linear, trace decreasing and completely positive
map $\mathcal{E}$. Trace decreasing means that
$\mathrm{tr}(\mathcal{E}(\rho)) \leq 1$ for all normalized density operators
$\rho$.  Complete positivity means that a map $\mathcal{E} \otimes \mathcal{I}$
is positive for all trivial extensions of a system (a system plus an
ancilla).  A quantum operation $\mathcal{E}$ can be represented (not
uniquely) in a form known as the Kraus representation or the
operator-sum representation
\begin{equation}
  \mathcal{E}(\rho) = \sum_{i} E_{i} \rho {E^{\dagger}_{i}},
  \label{1}
\end{equation} 
where the Kraus operators or operation elements $\{E_i\}$ satisfy the
following condition
\begin{equation}
  \sum_i E^\dagger_i E_i \leq I.
\end{equation}
Moreover, a map $\mathcal{E}$ which can be written in the form
(\ref{1}) is a quantum operation.  If the Kraus representation
contains only one operator we say that a quantum operation
$\mathcal{E}$ is pure.  The unitary evolution of density operator is
an example of a pure quantum operation.  Note that for a trace
preserving operation the Kraus operators $\{E_i\}$ satisfy a
completeness relation $\sum_i E^\dagger_i E_i = I$.  The von Neumann{\textendash}L{\"u}ders
measurement with no selection is an example of a trace preserving
quantum operation.  In this paper, we show that a destruction of
states \cite{CRSW} in a one-qudit system is a quantum operation by
finding its Kraus representation.

Denote the Hilbert space of a system consisting of one-qudit (a
$d$-dimensional quantum system, with $d$ finite) by $\mathcal{H}^1 \cong
\mathbb{C}^d$.  A state of the system is described by a density
operator $\rho \in \mathrm{End}(\mathcal{H}^1)$, where
$\mathrm{End}(\mathcal{H}^1)$ is the endomorphism space of
$\mathcal{H}^1$.  Such a description is not sufficient if a qudit can
be destroyed by a measurement apparatus.  This issue can be easily
solved by introducing the vacuum state $\left|\mathrm{vac}\right>$,
which is orthogonal to any vector from $\mathcal{H}^1$ and for all
observables $\hat{\Lambda}$ acting on $\mathcal{H}^1$ we adopt that
$\hat{\Lambda}|\mathrm{vac}\rangle = 0$.  The vector $|\mathrm{vac}\rangle$ spans the
one-dimensional vacuum space $\mathcal{H}^0 \cong \mathbb{C}$.  Therefore,
the Hilbert space $\mathcal{H}$ of the system under consideration is
the direct sum $\mathcal{H}^1 \oplus \mathcal{H}^0$, and the states are
mixtures of the elements from $\mathrm{End}(\mathcal{H}^1)$ and
$\mathrm{End}(\mathcal{H}^0)$.

We now briefly review the destruction of states procedure introduced
in Ref.\ \cite{CRSW}.  Let $\hat{\Lambda}$ be an arbitrary observable on
$\mathcal{H}^1$ with the spectrum $\Lambda$ and $\Omega$ be a subset of $\Lambda$.  We
denote the subspace spanned by all the eigenvectors corresponding to
the eigenvalues from the subset $\Omega$ by $\mathcal{H}_\Omega \subset \mathcal{H}^1$
and the projector onto this subspace by $\Pi_\Omega$.  If the qudit state is
an element of $\mathrm{End}(\mathcal{H}_\Omega)$ then the qudit is
destroyed, otherwise it is not.  First, we measure with no selection
the observable $\Pi_\Omega$.  The density operator $\rho \in
\mathrm{End}(\mathcal{H}^1) \oplus \mathrm{End}(\mathcal{H}^0)$ is reduced
to $\Pi_\Omega^\perp \rho \Pi_\Omega^\perp + \Pi_\Omega \rho \Pi_\Omega$, where $\Pi_\Omega^\perp=\mathrm{id}_{\mathcal{H}}
- \Pi_\Omega$.  Then, after the destruction we get either the vacuum with the
probability $\mathrm{tr}(\rho \Pi_\Omega)$ or the one-qudit state with the
probability $\mathrm{tr}(\rho \Pi_\Omega^\perp)$.  Therefore, the density operator
is reduced to $\Pi_\Omega^\perp \rho \Pi_\Omega^\perp + \mathrm{tr}(\rho \Pi_\Omega) |\mathrm{vac}\rangle
\langle\mathrm{vac}|$.  Formally, the destruction with no selection in the
set $\Omega$ of a one-qudit state $\rho$ is defined by the Kraus map $D_\Omega(\rho)\colon
\mathrm{End}(\mathcal{H}^1) \oplus \mathrm{End}(\mathcal{H}^0) \to
\mathrm{End}(\mathcal{H}^1) \oplus \mathrm{End}(\mathcal{H}^0)$, such that
\begin{equation}  
  D_\Omega(\rho)= \Pi_\Omega^\perp \rho \Pi_\Omega^\perp +
  \widehat{\mathrm{tr}}(\Pi_\Omega \rho \Pi_\Omega), 
  \label{destrukcja}
\end{equation}
where the supertrace $\widehat{\mathrm{tr}}$ (we call
$\widehat{\mathrm{tr}}$ a supertrace because it is a superoperator,
i.e., it is the operator in the endomorphism space{\textemdash}see e.g.,
\cite{Pres,Cav99}) is a linear map $\widehat{\mathrm{tr}}\colon
\mathrm{End}(\mathcal{H}) \to \mathrm{End}(\mathcal{H}^0)$ such that its
action on the endomorphism of the form $|\chi\rangle \langle\phi| \in
\mathrm{End}(\mathcal{H})$ is defined as follows
\begin{equation}
  \widehat{\mathrm{tr}}(|\chi\rangle \langle\phi|)=
  \langle \phi | \chi \rangle |\mathrm{vac}\rangle \langle\mathrm{vac}|.
  \label{superslad}
\end{equation}
The supertrace is a quantum operation
\begin{align}
  \widehat{\mathrm{tr}}(|\chi\rangle\langle\phi|) & = \sum_{i=0}^d \langle\phi|i\rangle \langle i|\chi\rangle
  |\mathrm{vac}\rangle \langle\mathrm{vac}| \nonumber\\
  & = \sum_{i=0}^d |\mathrm{vac}\rangle \langle i|\chi\rangle \langle\phi|i\rangle\langle\mathrm{vac}|
  \label{operacja superslad}
\end{align}
with operation elements $F_i = |\mathrm{vac}\rangle \langle i|$, where
$\{|i\rangle\}_{i=0}^{d-1}$ is an orthonormal basis of $\mathcal{H}^1$, and
$|d\rangle \equiv |\mathrm{vac}\rangle$.  Note that the quantum operation
(\ref{operacja superslad}) preserves the trace
\begin{equation}
  \sum_{i=0}^d F_i^\dagger F_i = \sum_{i=0}^d |i\rangle
  \langle\mathrm{vac}|\mathrm{vac}\rangle\langle i|= \sum_{i=0}^d |i\rangle\langle i| 
  = \mathrm{id}_{\mathcal{H}}.
\end{equation}
For simplicity we assume that the basis in $\mathcal{H}_\Omega$ is a subset
of $\{|i\rangle\}_{i=0}^{d-1}$.  Therefore, $D_{\Omega}(\rho)$ may be written as
\begin{equation}
  D_\Omega(\rho) = \Pi_\Omega^\perp \rho {\Pi_\Omega^\perp}^\dagger + \sum_{|i\rangle \in \mathcal{H}_\Omega} (F_i \Pi_\Omega) \rho
  {(F_i \Pi_\Omega)}^{\dagger}.
  \label{reprezentacja Krausa}
\end{equation}
Thus, $D_{\Omega}$ is also a quantum operation with operation elements
\begin{subequations}
\begin{gather}
  E_i = F_i \Pi_\Omega, \quad i = 0,\ldots, d, \\ 
  E_{d+1} = \Pi_\Omega^{\dag}.
\end{gather}
\end{subequations}
Note that $E_i = 0$ for $|i\rangle \notin \mathcal{H}_\Omega$.  Moreover, $D_\Omega$ is a
trace preserving quantum operation
\begin{align}
  \sum_{k=0}^{d+1} E_k^\dagger E_k & = 
  {\Pi_\Omega^\perp}^\dagger \Pi_\Omega^\perp + \sum_{|i\rangle \in \mathcal{H}_\Omega} \Pi_\Omega^\dagger 
  F_i^\dagger F_i \Pi_\Omega \nonumber\\
  & = \Pi_\Omega^\perp + \Pi_\Omega = \mathrm{id}_{\mathcal{H}}.
\end{align} 

As an example, consider destruction of states for one qubit ($d = 2$).
Let $\{|0\rangle, |1\rangle\}$ be an orthonormal basis of $\mathcal{H}^1$.  Assume
that $\hat{\Lambda} |0\rangle = |0\rangle$ and $\hat{\Lambda} |1\rangle = -|1\rangle$, so $\Lambda = \{-1, 1\}$.
Therefore, we have the following cases:

(i) $\Omega = \{1\}$, the destruction with no selection takes place if the
qubit is in the state $|0\rangle$.  It means that $\Pi_\Omega = |0 \rangle \langle 0|$ and
$\Pi_\Omega^\perp = |1 \rangle \langle 1| + |\mathrm{vac}\rangle \langle\mathrm{vac}|$.  In this case
operation elements for $D_\Omega$ are given by
\begin{subequations}
\begin{align}
  E_0 & = |\mathrm{vac}\rangle \langle 0|,\\
  E_3 & = |1\rangle \langle 1| + |\mathrm{vac}\rangle \langle\mathrm{vac}|.
\end{align}
\end{subequations}

(ii) $\Omega = \{-1\}$, the destruction with no selection takes place if the
qubit is in the state $|1\rangle$.  It means that $\Pi_\Omega = |1 \rangle \langle 1|$ and
$\Pi_\Omega^\perp=|0 \rangle \langle 0| + |\mathrm{vac}\rangle \langle\mathrm{vac}|$.  In this case
operation elements for $D_\Omega$ are given by
\begin{subequations}
\begin{align}
  E_1 & = |\mathrm{vac}\rangle \langle 1|,\\
  E_3 & = |0\rangle \langle 0| + |\mathrm{vac}\rangle \langle\mathrm{vac}|.
\end{align}
\end{subequations}

(iii) $\Omega = \{1, -1\}$, the destruction with no selection takes place if
the qubit is either in the state $|0\rangle$ or in the state $|1\rangle$.  It
means that $\Pi_\Omega = |0 \rangle \langle 0| + |1 \rangle \langle 1|$ and $\Pi_\Omega^\perp = |\mathrm{vac}\rangle
\langle\mathrm{vac}|$.  In this case operation elements for $D_{\Omega }$ are
given by
\begin{subequations}
\begin{align}
  E_0 & = |\mathrm{vac}\rangle \langle 0|,\\ 
  E_1 & = |\mathrm{vac}\rangle \langle 1|,\\
  E_3 & = |\mathrm{vac}\rangle \langle\mathrm{vac}|.
\end{align}
\end{subequations}

In conclusion, we have shown that destruction of states for one qudit
is a trace preserving quantum operation by finding explicitly its
Kraus representation.  This result seems to be useful in the study of
noisy quantum communication channels.

We are grateful to J. Rembieli{\'n}ski for useful discussions.  This work
was supported by University of {\L}{\'o}d{\'z} and the Laboratory of Physical
Bases of Processing of Information (\mbox{LFPPI}).


\end{document}